# A Foundational Approach to Physics

## Peter Rowlands

*IQ Group and Science Communication Unit, Department of Physics, University of Liverpool, Oliver Lodge Laboratory, Oxford Street, P.O. Box 147, Liverpool, L69 7ZE, UK. e-mail prowl@hep.ph.liv.ac.uk and prowl@csc.liv.uk*

*Abstract.* A core level of basic information for physics is identified, based on an analysis of the characteristics of the parameters space, time, mass and charge. At this level, it is found that certain symmetries operate, which can be used to explain certain physical facts and even to derive new mathematical theorems. Applications are made to classical mechanics, electromagnetic theory and quantum mechanics.

**1 A foundational level**

Certain aspects of physics suggest the existence of a core level of basic information which is completely independent of any hypothesis or model-building. This information is concerned with the definition of the fundamental parameters of measurement and how they are structured. Here, we are not so much describing nature itself as specifying the characteristics of the simplest categories needed to make such a description possible. It is, of course, sometimes argued that physics should not necessarily be concerned with the simplest possible ideas because nature may well not be simple in principle, but this argument is based on a misconception. Physics as we know it has evolved because it has created a set of simple categories which have been successful in devising the human construct that we call the 'description of nature'. Whether or not such simplicity is truly characteristic of 'reality' is a question outside the realm of physics.

    Now, the purpose of isolating a foundational level for physics would be to give a simple account of those important facts which are purely concerned with our own processes of measurement, and not with, say, the nature of matter or the structure of the universe. Such information, if isolated, could be dealt with more efficiently than if linked, unnecessarily, with more specific or more complicated theories, and could lead to the creation of wholly new sources of foundational information. In addition, if we pitch our foundational theories at a high level of sophistication, as we are often tempted to do, we cut ourselves off from understanding their origins. Many truly fundamental results have often turned out to be simple in principle, even when it has taken a sophisticated approach to find them. The simple bases are often found only after a prolonged struggle with more complicated ideas, but it is possible that we could discover some of them more easily by a direct analysis.

In attempting to reach the foundational level, then, we need to separate out the truly fundamental ideas from the mass of sophistications which inevitably accompany them, but the really basic ideas may not be particularly difficult to find. Obviously, for example, space and time are basic; it is impossible to conceive a theory of physics without space and time. This is not necessarily true for a combined space-time; combinations, by definition, are not basic. If we use space-time as basic, we lose sight of the separate identities of space and time, which are surely important at the fundamental level. It is certainly not true that in *combining* them we have explained one in terms of the other. We should really regard space-time, whether curved or otherwise, as a sophistication, as something to be discovered after we have investigated the separate identities of space and time.

Once we have selected space and time, there is only one other type of information that is likely to fit the description 'basic', for the fundamental bases of the whole of physics are undoubtedly the four known interactions: gravity; electromagnetism; and the strong and weak nuclear forces. Once we truly understand these, then we will also understand physics. But we already know something about them. We know, for example, that the *source terms* for gravity and the electromagnetic interaction are, respectively, mass and electric charge. We don't know the source terms for the weak and strong forces as precisely, but we know that such terms must exist; and, according to the rules of quantum electrodynamics, they should be more like electric charge than like mass. Also, although the three nongravitational forces are very different under normal conditions, the Grand Unified Theories of particle physics suggest that, under ideal conditions, they would be identical. It is reasonable, therefore, to assume that the differences between these forces are not basic, but a 'sophistication', to be explained in terms of the physical particles that actually exist, after we have established the basics. I have, therefore, found it convenient to describe the unknown sources for the weak and strong interactions as weak and strong 'charges', and to refer sometimes to the three nongravitational sources under the collective label 'charge'. Something like this concept is actually used when we talk about the process of 'charge conjugation'.

Thus, we have four basic parameters – space, time, mass and charge – and we can use such techniques as dimensional analysis to show that all other physical parameters arise from compounded versions of these elementary ones. We can also show that it is precisely these parameters which are assumed to be the elementary ones in the statement of the CPT theorem. It is perhaps slightly surprising that physicists have been so often prepared to tackle fundamental questions without taking proper account of these basic ideas. Although we can't hope to *analyse* really basic ideas, we can learn a great deal by setting one off against the other. It would surely be profitable to examine the properties of these parameters as closely as possible and look for patterns, or symmetries of one sort or another, that would help to clarify their meaning and uses. Symmetry has been such a powerful tool in understanding particle physics and the fundamental interactions that we have every reason to expect to find it here. We should, at any rate, examine the properties of our parameters to see whether or not it exists.

## 2 Conserved and nonconserved: mass and charge versus space and time

Perhaps the first thing that we notice when we closely examine space, time, mass and charge is that the last two are conserved quantities and the first two nonconserved. The conservation laws of mass and electric charge are among the most fundamental in physics, and we have every reason to believe that they are true without exception. Almost certainly, also, some type of conservation law applies to the other two types of charge and manifests itself in such properties of fundamental particles as lepton and baryon conservation. In addition, the conservation laws of mass and charge are not merely global, applying to the total amount of each quantity in the universe, but also *local*, applying to the amount of each quantity at a given place in a given time. It is as though each element of mass or charge had an *identity* which it retained throughout all interactions, subject only, in the case of charge, to its annihilation by an element with the opposite sign. We could, in principle, label each unique element with an identity tag which it would never lose. Mass and charge, thus, have identical conservation properties, apart from the fact that masses have no elements with opposite sign.

When we look at nonconservation, as manifested by space and time, we might at first imagine that it is merely the absence of conservation, but this is not so. Examination shows that it is the *exact opposite* of conservation and it is just as definite a property. Nonconservation is, in fact, one of the most interesting and important of all physical properties, and it manifests itself in many different ways. Thus, just as the elements of mass and charge have individual, specific and permanent identities, so those of space and time have no identity whatsoever, and this fact has to be incorporated directly into physics at all levels. We refer, for example, to the property of *translation symmetry* for both space and time. This means that every element of space and time is exactly like every other, and is not only indistinguishable in practice, but *must be stated to be indistinguishable* when we write down physical equations. And this translation symmetry is not an insignificant philosophical concept; it is responsible for two of the most important laws in nature, for Noether's theorem shows us that the translation symmetry of time is precisely identical to the conservation of energy, and that the translation symmetry of space is precisely identical to the conservation of linear momentum.

Space, in addition, because it is three-dimensional, also has rotation symmetry; this means that there is no identity, either, for spatial *directions*. In addition to having no unique elements, space also lacks a unique set of dimensions. One direction in space is identical to any other; this is the fact that is responsible for space's *affine* structure, the infinite number of possible resolutions of a vector into dimensional components. Space rotation symmetry is a very important property and we will return to it later. Noether's theorem shows that it is exactly the same thing as the conservation of angular momentum.

The exactly opposite nature of conservation and nonconservation could be illustrated by expressing the identity or uniqueness properties of mass and charge in terms of 'translation' *a*symmetries. Translation asymmetry then means that one element of mass or charge cannot be 'translated to' (or exchanged for) any other within a system, however similar.

But translation and rotation symmetry are not the only manifestations of nonconservation in space and time. The whole of physics is based on defining systems in which conserved quantities remain fixed while nonconserved quantities vary absolutely. A conserved quantity can only be defined with respect to changes in a nonconserved quantity. In effect, we look at how mass and charge, and such quantities as energy, momentum, force or action remain constant, or zero, or a maximum or a minimum, because of the more fundamental requirements involving mass and charge, while the space and time coordinates alter arbitrarily. The alteration of space and time is expressed by describing them in terms of differentials. The very fact that we have based physics on differential equations and the definition of systems involving conservation requirements is an expression of the presence of both absolutely conserved and absolutely nonconserved terms in nature.

The absoluteness of the nonconservation properties is manifested in the *gauge invariance* used in both classical and quantum physics. In classical or quantum electrodynamics, electric and magnetic fields terms remain invariant under arbitrary changes in the vector and scalar potentials, or phase changes in the quantum mechanical wavefunction, brought about, essentially, by translations (or rotations) in the space and time coordinates. Gauge invariance tells us, in effect, that a system will remain conservative under arbitrary changes in the coordinates which do not produce changes in the values of conserved quantities such as charge, energy, momentum and angular momentum. In other words, we cannot know the absolute phase or value of potential because we cannot choose to fix values of coordinates which are subject to absolute and arbitrary change. Even more significantly, in the Yang-Mills principle used in particle physics, the arbitrary phase changes are specifically *local*, rather than global. Nonconservation, therefore, must be local in exactly the same way as conservation.

**3 Real and imaginary: space and mass versus time and charge**

Now, space and time are alike in their nonconservation, but we know that there must be fundamental differences between them; otherwise, they would be indistinguishable. One such distinction is evident in the very mathematical combination which produces four-dimensional space-time. This is the fact that, while Pythagorean addition produces positive values for the squares of the three spatial dimensions, the squared value of time becomes negative. A convenient way to represent time, then, is by an imaginary number, as in the Minkowski space-time 4-vector used in relativity. This, of course, does not make time 'imaginary' in itself; but it is important for us to ask why this particular 'trick' actually works. It is not really adequate to describe it as a 'convenience' without explaining why it is convenient. One interesting fact is that an imaginary representation would also make uniform velocity imaginary, while acceleration would remain real.

To try to get beyond the facile explanation that the trick is good because it works, we should see if we can learn anything relevant from the representation of mass and charge. Here, we have the intriguing fact, long known but never explained, that forces between like masses are attractive, whereas forces between like (electric) charges are repulsive; that is, the forces between like masses and like charges have opposite signs. Now, the force laws effectively square mass and charge terms, in the same way as space and time terms are squared in Pythagorean addition. Suppose, then, that we choose to represent charges by imaginary numbers and masses by real ones. We then have a symmetrical representation for the Newton and Coulomb force laws:

$$F = -\frac{Gm_1 m_2}{r^2}$$

$$F = -\frac{iq_1 iq_2}{4\pi\varepsilon_0 r^2}$$

In addition, of course, the other two forces – the strong and weak interactions – are like the electromagnetic in being repulsive for like particles, and so the source terms for these forces would also presumably be defined by imaginary numbers. But the three types of source would have to be distinguished from each other in some way. And here we have a stroke of luck, for the mathematics required for such a situation is already available and has been well-known for a hundred and fifty years. This is the *quaternion* system, discovered by Hamilton in 1843, in which $\boldsymbol{i}, \boldsymbol{j}$ and $\boldsymbol{k}$, the three square roots of $-1$, are related by the formulae:

$$\boldsymbol{i}^2 = \boldsymbol{j}^2 = \boldsymbol{k}^2 = \boldsymbol{ijk} = -1 \ .$$

For historical reasons, quaternions became a proscribed concept at the end of the nineteenth century, and they still have a reputation for being 'difficult' or 'esoteric'; but, in fact, they are remarkably easy to use, being effectively the reverse of the 4-vectors used in Minkowski space-time: three imaginary parts and one real (ordinary real numbers), as opposed to three real parts and one imaginary. But the real significance of quaternions is that they are unique. As Frobenius proved in 1878, no other extension of ordinary complex algebra involving imaginary dimensions is possible: if we require a dimensional imaginary algebra (as the source terms for the electromagnetic, strong and weak interactions suggest we might) then we have only one possible choice – an algebra based on one real part and three imaginary. (This is, of course, if we wish to retain associativity. One further extension exists in the 8-part octonions or Cayley numbers, which break associativity, and which are discussed later in the paper.)

Hamilton discovered the quaternions after finding that a system with two imaginary parts was impossible, and, almost immediately, he felt that he was on to the true explanation of 3-dimensional space, with time taking up the fourth or real part. By our analysis, it would be more convenient to apply them to the three imaginary components of charge, with mass taking up the real part. However, space and time would then become a three real- and one imaginary-part system by *symmetry*. In this sense, the three components of charge (say, $\boldsymbol{i}e, \boldsymbol{j}s, \boldsymbol{k}w$) could be considered as the 'dimensions' of a single charge parameter, with their squared values used in the calculation of forces added, in the same way as the three parts of space, by Pythagorean addition:

| | | | | |
|---|---|---|---|---|
| space-time | **i**$x$ | **j**$y$ | **k**$z$ | $it$ |
| mass-charge | **i**$e$ | **j**$s$ | **k**$w$ | $m$ |

One remarkable consequence of adopting this symmetry as an exact one would be that the vector property of space would be extended to incorporate a quaternionic-like 'full' product between two vectors, combining the scalar product with $i$ times the vector product. It is the extra vector terms in this product which are responsible for the otherwise 'mysterious' spin property in quantum mechanics.

It is here that we can now return to the subject of the rotation symmetry of space. If charge, like space, is a three-dimensional parameter, then we need to investigate how the dimensions behave with respect to each other. Immediately, we should expect a difference from space, since charge is a conserved quantity. In fact, we should expect conservation in dimension as well as in quantity; in principle, charge should exhibit rotation *a*symmetry. That is, the sources of the electromagnetic, weak and strong interactions should be separately conserved, and incapable of interconversion. Immediately, this should tell us that the proton, which has a strong charge measured by its baryon number, cannot decay to products like the positron and neutral pion, which have none. Attention to basics here would require the separate conservation of the three charges to be built into Grand Unified Theories. Particle theorists have been puzzled as to why the proton does not decay; but basic reasoning suggests that there may be an answer. (The Weinberg-Salam unification of electromagnetic and weak forces is not, of course, affected because this theory is a statement of the identity of effect in the two interactions, under ideal conditions, not of identity of the sources; the three quaternion operators *i*, *j* and *k* are different sources, though identical in effect.) In addition, separate conservation laws would easily lead to baryon and lepton conservation, baryons being the only particles with strong, as well as weak, components, and leptons being the only particles with weak, but no strong, components.

In view of such advantages in applying an imaginary representation to the three types of charge, we may be inclined to ask if there is any further benefit; and, in fact, there is, for imaginary numbers have yet another important property. This is the fact that equal representation must be given to positive and negative values of imaginary quantities. Unlike real numbers, imaginary ones allow neither positive nor negative values to be privileged in algebraic equations. In other words, every equation which has a positive solution also has an algebraically indistinguishable negative solution (the complex conjugate). Thus, all our charges (but not necessarily masses) must exist in both positive and negative states. This is the precise requirement for the existence of antiparticles; even those particles, such as the neutron and neutrino, which have no electric charge still have antiparticles because they have strong and/or weak charges whose signs may be changed (under the process of charge conjugation, already mentioned).

**4 Divisible and indivisible: space and charge versus time and mass**

Now space is like time in being nonconserved, like mass in being real, and, apparently, like charge in being dimensional. Dimensionality, however, doesn't really look like a basic property. Is there any basic property which explains it? It seems very probable that there is, and, in looking at this question, it will be necessary once more to examine the relationship between space and time.

Space and time have often been assumed to be alike in most respects, but there is good evidence that they are fundamentally different. Space, for example, is always used in direct measurement; in fact, it is impossible to measure anything but space. Our 'time'-measuring devices, such as pendulums, mechanical clocks, and crystal and atomic oscillators, all use some concept of repetition of a spatial interval. Special conditions have to be used to set up such measurements, whereas any object whatsoever can be used to measure space. Space also is reversible – and it is this reversibility which is used in the measurement of time – but time is not.

It may be convenient here to mention the famous paradoxes of Zeno of Elea. Of course, these are, in a sense, 'answered' by the use of limits or infinite series, but Whitrow, who has made the most extensive and influential recent study, thinks that the answers still leave the problem incomplete.[1] In the well-known argument about the race between Achilles and the Tortoise, Achilles, in any number of time intervals, should never catch up with the Tortoise, to whom he has given a lead, because, each time he thinks he has caught up, he finds the Tortoise has already moved further ahead, even if only by an ever smaller amount. Another example is the Dichotomy Paradox, in which an object moving over any distance can never get started because it must cover half the distance before it covers the whole, and a quarter of the distance before it covers half, and so on; to go any distance in a finite amount of time, it must already have been involved in an infinite number of operations.

The problem seems to be the infinite divisibility of time; Achilles, for example, never catches the Tortoise because we have assumed that the time for the race can be divided up into finite intervals. On the basis of these, and similar paradoxes, Whitrow writes: 'One can, therefore, conclude that the idea of the infinite divisibility of time must be rejected, or ... one must recognize that it is ... a logical fiction.' And the more recent authors, Peter Coveney and Roger Highfield conclude that: 'Either one can seek to deny the notion of 'becoming', in which case time assumes essentially space-like properties; or one must reject the asumption that time, like space, is infinitely divisible into ever smaller portions.'[2] The paradoxes seem to show, according to Whitrow, that motion is 'impossible if time (and, correlatively, space) is divisible ad infinitum'[5]

Zeno's paradoxes pose fundamental problems for the nature of space and time; Bertrand Russell considered them 'immeasurably subtle and profound', and Alfred North. Whitehead thought that they showed an 'instant of time' to be 'nonsense'.[2] Our reason for including them here is to show that there is good evidence that one cannot simply assume that time can be indefinitely subdivided like space. There is every reason to believe, in fact, that time, unlike space, is an absolute continuum. There is no infinite succession of measurable instants in time, as supposed in the paradoxes, because there are no instants. Time cannot actually be divided. To use a more contemporary jargon, space is digital, time is analogue – and we have both concepts in nature because we have both parameters.

We can also say that time is the set of reals with the standard topology superimposed (and is nonalgorithmic); space is the set of reals without the topology (and is algorithmic). Henri Bergson, according to Whitrow, 'enthusiastically adopted the view' that time 'is wholly indivisible', 'as a means of escaping the difficulties raised by Zeno, concerning both temporal continuity and atomicity, without abandoning belief in the reality of time. ... Unfortunately, in attacking the geometrization (or spatialization) of time he went too far and argued that, because time is essentially different from space, therefore it is fundamentally irreducible to mathematical terms.'

Continuity is a word with many meanings, and different uses of the word have caused confusion. The 'continuity' attributed to space because of its indefinite divisibility is not what is meant by the absolute continuity of time. Absolute continuity cannot be visualised and any process used to describe it would deny continuity. The property which space has that is often referred to as 'continuity' is indefinite elasticity, its 'continual' recountability or its unending divisibility. But the very divisibility of space is what denies it *absolute* continuity; and the elastic nature of the divisibility comes from the entirely different property of nonconservation. We expect a nonconserved quantity to have nonfixed units, but they are units nonetheless. The whole process of measurement depends crucially on the divisibility of space, or creation of discontinuities within it. Thus the entire problem of Zeno's paradoxes disappears as soon as we accept that we can have discontinuities or divisibility in space, but not in time.

Space can be discontinuous in both quantity and direction; it can be reversed and changed in orientation; and, without both of these properties, measurement would be impossible. Time, however, cannot be reversed, precisely because it is absolutely continuous. Any reversal of time would require discontinuity. For the same reason, time cannot be multidimensional, or, in our terminology, 'dimensional'. The same distinction occurs between mass and charge. Mass is an absolute continuum present in all systems and at every point in space (if only in the form of fields and energy); this is why there is no negative mass, for negative mass would necessarily require a break in the continuum. Charge, on the other hand, is divisible and observed in units; of course, because charge is a conserved quantity, unlike space, these units must be fixed, unlike those of space. Again, charge as a noncontinuous quantity is also dimensional, and, thus we might suggest divisibility as the 'cause' of dimensionality; and, though it is not a direct argument that divisibility causes dimensionality, it is immediately apparent why absolutely continuous quantities must be *non*dimensional.

There is, however, a more direct argument for the dimensionality of discrete quantities. One cannot, in fact, demonstrate discreteness in a one-dimensional system. Though we think of a line as one-dimensional, it is in fact no such thing: it is a one-dimensional construction within a two-dimensional one. If our space was truly one-dimensional we would only have a point with no extension. We couldn't demonstrate discreteness, and certainly not discreteness with variability, as we demand of space. Interestingly, it is dimensionality as such, rather than any particular level of dimensionality which is responsible for creating the additional level of discreteness required by the introduction of algebraic numbers, and even of transcendental numbers such as $\pi$; two independent dimensions are sufficient to create the required level of incommensurability at the rational number level, and the introduction of a third dimension requires no qualitatively new type of number.

To return to time's nondimensionality, one often reads about a 'reversibility paradox', where time, according to the laws of physics is reversible in mathematical sign, when it is clearly not reversible in physical consequences. Time, however, we need to remember, is characterised by imaginary numbers, and imaginary numbers are not privileged according to sign. Thus, it is quite possible to have a time which has equal positive and negative mathematical solutions because it is imaginary, but which has only one physical direction because it is continuous. (The corresponding unipolarity, or single sign, of mass is the reason why we have a CPT, rather than an MCPT, theorem, C standing for charge conjugation, P for space reflection and T for time reversal, all of which have two mathematical sign options.)

The distinction between space and time has many interesting consequences. In principle, when we mathematically combine space and time in Minkowski's 4-vector, as symmetry apparently requires us to do, we have two options: we can either make time space-like (or discrete) or space time-like (or continuous). This seems to be the origin of wave-particle duality. The discrete options lead to particles, special relativity and Heisenberg's quantum mechanics.[4,5,6] The continuous options lead to waves, Lorentzian relativity and Schrödinger's wave mechanics. Heisenberg makes everything discrete, so mass becomes charge-like quanta in quantum mechanics; Schrödinger, on the other hand, makes everything continuous, so charge becomes mass-like wavefunctions in wave mechanics. In measurement, the true situations are restored, for Heisenberg reintroduces continuous mass via the uncertainty principle and the virtual vacuum, while Schrödinger reintroduces discrete charge via the collapse of the wavefunction. (With the altered parameters represented by S*, T*, M*, C*, the respective options are S, T*, M*, C (Heisenberg) and S*, T, M, C* (Schrödinger).)

Another aspect of the distinction between space and time occurs in the fundamental fact that time, in the definition of velocity and acceleration, the basic quantities used in dynamics, is the independent variable, whereas space is the dependent variable. This situation arises because time, unlike space, is not susceptible to measurement. We have no control over the variation of time, and so its variation is necessarily independent.

The fundamental distinction between the status of space and time almost certainly also has relevance in mathematics. In the seventeenth century, there were two processes of differentiation: the discrete (or Leibnizian), essentially modelled on variation in space; and the continuous (or Newtonian), essentially modelled on variation in time. Like particles and waves, each is a valid option, for differentiation is a property linked to nonconservation, and not concerned, in principle, with the difference between absolute continuity and indefinite divisibility. (The solutions of Zeno's paradoxes that invoke the concept of limit tacitly assume the Newtonian definition of differentiation.) Again, it is probable that the Cantorian definition of an absolutely continuous set of real numbers has equal validity with the idea of an infinitely constructible, though not absolutely continuous, set of real numbers based on algorithmic processes. (We may note here the fundamental significance of the Löwenheim-Skolem theorem, that any consistent finite, formal theory has a denumerable model, with the elements of its domain in a one-to-one correspondence with the positive integers.) The mathematical options that are available, here and elsewhere, are almost certainly a reflection of the availability of physical options. Continuity and discontinuity, finiteness and infinity, and so on, probably exist as mathematical categories because they are also physical categories.

Abraham Robinson, in his *Non-Standard Analysis*,[3] treats infinitesimals as though they had the properties of real numbers, and says that proofs of many theorems become much simpler by this method, although all non-standard proofs may be duplicated by standard ones (and vice versa). Non-standard analysis can also be related to Skolem's non-standard arithmetic of 1934, with its denumerable model of the reals, and what has been described as non-Archimedean geometry, which relates this to space. These versions of non-standard mathematics are a reflection of the discrete nature of space while 'standard' results (based on limits) rely on the continuity of time.

## 5 A group of order 4

From what we have seen, then, the four basic parameters seem to be distributed between three sets of opposing paired categories: real / imaginary, conserved / nonconserved, divisible / indivisible, with each parameter paired off with a different partner in each of the categories, according to the following scheme:

| | | | |
|---|---|---|---|
| **space** | real | nonconserved | divisible |
| **time** | imaginary | nonconserved | indivisible |
| **mass** | real | conserved | indivisible |
| **charge** | imaginary | conserved | divisible |

The properties where they match, seem to be exactly identical, and where they oppose, to be in exact opposition. (Certain representations, however, like the Dirac equation involve mathematical reversals of physical properties, the Lorentz-invariant structure demanding either timelike space or spacelike time, with corresponding reversals in the properties of mass or charge.) Mathematically, this scheme incorporates a group of order 4, in which any parameter can be the identity element and each is its own inverse.[7,8,9]

An algebraic representation is easily accomplished by representing the properties of space (real, nonconserved, divisible) by, say, *a*, *b*, *c*, with the opposing properties (imaginary, conserved, indivisible) represented by –*a*, –*b*, –*c*. The group now becomes:

| | | | |
|---|---|---|---|
| **space** | *a* | *b* | *c* |
| **time** | –*a* | *b* | –*c* |
| **mass** | *a* | –*b* | –*c* |
| **charge** | –*a* | –*b* | *c* |

With group multiplication rules of the form:
$$a * a = -a * -a = a$$
$$a * -a = -a * a = -a$$
$$a * b = a * -b = 0$$

and similarly for *b* and *c*, we can establish a group multiplication table of the form:

| * | space | time | mass | charge |
|---|---|---|---|---|
| space | space | time | mass | charge |
| time | time | space | charge | mass |
| mass | mass | charge | space | time |
| charge | charge | mass | time | space |

This is the characteristic multiplication table of the Klein-4 group, with space as the identity element and each element its own inverse. However, there is no reason to privilege space with respect to the other parameters, since the symbols *a* and –*a*, *b* and –*b*, *c* and –*c* are arbitrarily selected, and any of the other three parameters may be made the identity by defining its properties as *a*, *b*, *c*. For example, if mass is made the identity element, then the group properties may be represented by:

| | | | |
|---|---|---|---|
| **space** | *a* | –*b* | –*c* |
| **time** | –*a* | –*b* | *c* |
| **mass** | *a* | *b* | *c* |
| **charge** | –*a* | *b* | –*c* |

and the multiplication table becomes:

| * | mass | charge | time | space |
|---|---|---|---|---|
| mass | mass | time | charge | space |
| charge | time | mass | space | charge |
| time | charge | space | mass | time |
| space | space | time | charge | mass |

Various further representations are possible, and seem to be relevant, in particular, to the mathematical structure of the Dirac equation. For example, the identity element, say mass, could be represented by the scalar part of a quaternion (1) and the other three terms by the imaginary operators *i*, *j*, *k*, if we choose only the modular values, and ignore the + and – signs:

| * | 1M | *i*C | *j*T | *k*S |
|---|---|---|---|---|
| 1M | 1M | *i*C | *j*T | *k*S |
| *i*C | *i*C | (–)1M | *k*S | (–)*j*T |
| *j*T | *j*T | (–)*k*S | (–)1M | (–)*i*C |
| *k*S | *k*S | *j*T | *i*C | (–)1M |

With the + and – signs added, we would require the full (and now cyclic) quaternion group structure of eight components. It is important to recognise here that the quaternion operators are extrinsically derived and not an integral component of the parameters space, time, mass and charge. Though the addition of these operators creates a new group structure, this structure is a relation between new mathematical constructs and not between the parameters themselves; it also presupposes the validity of the original symmetry between the parameters.

If the 3-dimensionality of charge and space is directly involved, the overall structure would require a quaternion and a 4-vector within another overall quaternion-type arrangement. This could be accomplished using an octonion, with sixteen members (±1*m*, ±*is*, ±*je*, ±*kw*, ±*et*, ±*fx*, ±*gy*, ±*hz*), though this is no longer a group. The nonassociativity of the dimensional terms in the octonion extension seems to be lost within terms which effectively cancel each other out, and are of no physical significance.

If charge is taken as the identity element, and is represented by a scalar, the remaining structure for time, space and mass (and, implicitly, the energy, momentum and mass operators) becomes that of the Dirac algebra, and *SU*(5) or *U*(5). Such representations do not determine the properties of space, time, mass and charge. They exist because the group has four components, and can, therefore, be represented by a 4-component structure like quaternions, in which the link between elements is made by a binary operation (squaring); but the link between a group with four components and a 4-dimensional space-time or mass-charge may be in itself significant.

Using the postulated group as a working hypothesis, it becomes possible to explore possible constraints on the laws of physics, as a result of group properties (as is shown below). Another area to be investigated might be the way in which the relationship between the quaternion representation and the requirement of separate conservation for charges might affect the fundamental particle structures that are possible.[8,9,10]

**6 Scaling relations**

The group elements are required to be their own inverses, and to be each identities. In addition, the group multiplication rule (when all possible arrangements are taken into consideration) requires:

$$\text{charge} * \text{time} = \text{space} * \text{mass} .$$

A binary operation which makes this possible is the squared multiplication of units, such as already exists for space and time in the 4-vector combination and for mass and charge when they are combined in a quaternion. It is also inherent in the description of charge, time, space and mass as, respectively, quaternion (or possibly pseudovector), pseudoscalar, vector and scalar, that the units of their squared quantities must be comparable numerically. To create the necessary number of independent fundamental relationships, we need to define three scaling constants (or rather scaling parameters, since they need not be actually constant if they are known to vary according to some fixed rule). And since the system has inherent duality in making each quantity its own inverse, then we must define a relation between each quantity and the inverse of every other, for which one further scaling constant (or parameter) will suffice. (The existence of the binary operation of squaring within the parameter group seems to be linked to the same operation being responsible for the 4-dimensionality of space-time and mass-charge.)

The group relationship predicts that such fundamental constants must exist, while effectively ensuring that their individual values have no independent meaning. To relate these to familiar scales of measurement, we create them from combinations of the four historically-generated fundamental constants $G$, $c$, $h$ (or $\hbar$), $4\pi\varepsilon_0$. (Here, for convenience, we assume that 'charge' has the electromagnetic value, though this is not a necessary assumption, and a grand unified value could be used instead; the actual 'values' of the constants are not particularly significant – only the fact that some such scaling must exist.)

We can now express the scaling relations between the units of space ($r$), time ($t$), mass ($m$), and charge ($q$) as follows (with the equality sign being interpreted as meaning 'equivalence'):

$$r = ict \qquad (1)$$

$$r = \frac{G}{c^2} m \qquad (2)$$

$$iq = (4\pi\varepsilon_O G)^{1/2} m \qquad (3)$$

The respective imaginary and quaternion operators required by $t$ and $q$ are significant in determining the signs of their squared units. These operators are normally subsumed within the symbols $t$ and $q$, but here they are added for emphasis.

The further relations between any parameter and the inverse of any other can all be derived from:

$$m = \frac{h}{c^2} \frac{1}{it} \qquad (4)$$

This last result is the one that we recognise as being responsible for quantization of energy and other physical properties. Quantization could thus be said to be a result of the fact that each parameter is its own inverse. Quantization and duality of scale are aspects of the same phenomenon.

The four independent scaling constants in the above scheme become $c$, ($G / c^2$), $(4\pi\varepsilon_O G)^{1/2}$, and ($h / c^2$) (or $\hbar / c^2$). These are merely the scaling relations between the units of each quantity, but the presence of $c^2$ and $h$ informs us that these quantities are fundamental to physics, whether classical, relativistic or quantum. In principle, any term related to another by a scaling relation in a meaningful physical equation can be replaced by that term to produce another meaningful equation.

A significant aspect of the binary operation between parameters is the squaring of the units of each, or the multiplication of a unit of any parameter by an identically-valued unit of the same parameter. Now, units of mass and charge have individual identities, unlike those of space and time, and so the 'squaring' of their units becomes the multiplication of individual units, such as $m_1 m_2$ and $q_1 q_2$, and such 'squaring' must be a universal operation between any units of mass and charge, no individual unit being privileged. It will be convenient to give this process the name of 'interaction'. (It will be recognised that 'interaction' in this sense is nonlocal.)

**7 Constructed quantities**

The most fundamental laws of physics are essentially definitions and conservation laws. Classical mechanics, for example, is structured on only two fundamental requirements: the construction of a quantity involving conserved and nonconserved parameters (force, energy, momentum, action, Lagrangian, Hamiltonian) and the definition of its behaviour under variation of the variable components, that is, whether it is defined to be zero, invariant, or an extremum. Essentially, the laws of classical mechanics are set up to define what is meant by a conservative system.

Now, the key concepts in classical mechanics (as in other aspects of physics) are those which combine the minimum information necessary to distinguish the conserved and nonconserved parameters. Of the conserved quantities, mass is universal and never zero, and therefore must be present; charge, however, is local, and can take zero values. To specify the conservation or invariability of mass, we also need to specify the nonconservation or variability of space and time; hence, these parameters are included in differential form. A convenient way to define a system, therefore, would be the construction of a quantity containing mass and the differentials of space and time. The most immediately useful constructs then include $p = m\, dr / dit$ and $F = dp / dit$ (with time, most conveniently specified as the independent variable). The second quantity, as has been previously explained, has the advantage of producing a real rather than an imaginary construct. Now, space, of course, is really a vector (neglecting, for the moment, any 4-vector aspects); to incorporate this aspect, we may multipy both terms by the unit vector $\mathbf{r} / r$, to yield the familiar quantities, *momentum*,

$$\mathbf{p} = m\frac{d\mathbf{r}}{dit}$$

and *force*,

$$\mathbf{F} = \frac{d\mathbf{p}}{dit} .$$

(Imaginary and quaternion labels are retained here for emphasis but would not, of course, normally be used.)

The definitions of such quantities are, as yet, purely mathematical and convey no additional physical information. This can now be supplied, however, by using the scaling relations to find other quantities to which these defined ones can be related, while at the same time applying the conditions for conservation and nonconservation. This enables us to set up a system of equations for classical mechanics and electromagnetic theory.

**8 Classical mechanics**

From (2) and (3), remembering that each element of mass is unique, we may derive the expression

$$Gm_1 m_2 = h\frac{r}{it}$$

In differential form, under the specific conservation of mass elements,

$$Gm_1 m_2 = h\frac{dr}{dit} ,$$

from which

$$\frac{Gm_1 m_2}{c^2 it} = m\frac{dr}{dit} = p .$$

The mass term on the right hand side, of course, is a new mass unit, distinguishable from $m_1$ and $m_2$.

By differentiation, and a further substitution,

$$-\frac{Gm_1m_2}{c^2 i^2 t^2} = -\frac{Gm_1m_2}{r^2} = \frac{dp}{dit}.$$

Applying the unit vector, $\mathbf{r} / r$, this becomes

$$-\frac{Gm_1m_2}{r^3}\mathbf{r} = \frac{d\mathbf{p}}{dit},$$

which is a combination of Newton's law of gravitation and second law of motion, with the left hand side a new equivalent quantity for force, conventionally described as *gravitational force*.

Neither $m_1$ nor $m_2$ is, of course, privileged, and so the equation can also be written in the form:

$$-\frac{Gm_1m_1}{r^3}\mathbf{r} = \frac{d\mathbf{p}}{dit}$$

Interpreting the vectors $\mathbf{r}$ and $\mathbf{p}$ as directed from $m_1$ to $m_2$ means that reversing the mass terms produces reversed vectors, from $m_2$ to $m_1$, as required by Newton's third law of motion.

The equivalent case for charges defines Coulomb's law of electrostatics and introduces *electrostatic force* (with the opposite sign, and hence reversed vector, for identically valued charges):

$$\frac{q_1 q_2}{4\pi\varepsilon_0 r^3}\mathbf{r} = \frac{d\mathbf{p}}{dit}.$$

All the other significant and relations of classical mechanics, in any of its forms, can be derived now by purely mathematical manipulation. For example, nterpreting a 'system' to mean any combination of unit masses, the conservation of momentum follows by integration of the total force over time, and the conservation of angular momentum ($\mathbf{L} = \mathbf{r} \times \mathbf{p}$) from the fact that $d\mathbf{p} / dt$ in a conservative system is zero.

Direct manipulation of the scaling relations reveals that momentum terms are equivalent to $mc\mathbf{r} / r$, and that scalar terms of the form $Gm_1m_2 / r$ and $q_1q_2 / 4\pi\varepsilon_0 r$, which we may describe as *gravitational and electrostatic potential energies*, are equivalent to those of the form $mc^2$; in each of these cases, $m$ may be described as an 'equivalent mass'. Though these results normally emerge only from relativity theory, they are actually inherent in the structure from which classical mechanics must be derived.

Further results follow from on immediately from the mathematical definition of new concepts. Thus, defining velocity as $\mathbf{v} = d\mathbf{r} / dit$ and acceleration as $\mathbf{a} = d\mathbf{v} / dit$, and field intensity as $\mathbf{F} / m$, we have, in the case of constant mass, $\mathbf{F} = m\mathbf{a}$, and can define *gravitational field intensity* as

$$\mathbf{g} = -\frac{Gm}{r^3}\mathbf{r}$$

and *electrostatic field intensity* as

$$\mathbf{E} = -\frac{iq}{4\pi\varepsilon_0 r^3}\mathbf{r}.$$

From vector theory, we can show that, for the related *scalar potentials*, $\phi = -Gm/r$ and $\phi = -iq/4\pi\varepsilon_0 r$,

$$\mathbf{g} = -\nabla\phi$$

and

$$\mathbf{E} = -\nabla\phi,$$

and, also, that the respective force laws are equivalent to the Laplace equations

$$-\nabla^2\phi = \nabla\cdot\mathbf{g} = 0$$

and

$$-\nabla^2\phi = \nabla\cdot\mathbf{E} = 0,$$

in a space without sources, and to the Poisson equations,

$$-\nabla^2\phi = \nabla\cdot\mathbf{g} = 4\pi\rho G$$

and

$$-\nabla^2\phi = \nabla\cdot\mathbf{E} = \rho/\varepsilon_0,$$

in a space with them. None of this requires any new physical argument.

**9 Classical electromagnetic theory**

To replace nonrelativistic equations with relativistic ones, we simply replace all vector terms with 4-vectors, $\mathbf{r}$, for example, being replaced by ($\mathbf{r}$, $ict$). This procedure can be done, of course, with purely mechanical equations, to generate the standard results of special relativity, but it is particularly significant in classical electromagnetic theory, which follows on immediately from applying 4-vector terms to the definition of electrostatic force.

The significant fact here is that charge is *locally* conserved, and, hence, by a standard argument, the continuity equation,

$$\frac{\partial\rho}{\partial t} + \nabla\cdot\mathbf{j} = 0,$$

must apply, with $\rho$ defined as the charge density and $\mathbf{j} = \rho\mathbf{v}$ as the current density. The differential operator in this equation is a 4-vector, and so, recognisably, is the quantity with scalar and vector parts, $\rho$ and $\mathbf{j}/c$.

Now, the scalar part of this latter quantity appears in Poisson's equation,

$$-\nabla^2 \phi = \rho / \varepsilon_0,$$

which is the differential form of Coulomb's law, and so we should expect to find an equivalent vector part ($\mathbf{A} / c$) for $\phi$, and an equivalent scalar part ($-(1/c^2) \partial^2 / \partial t^2$) for $\nabla^2$. Since the new differential operator $\square = ((1/c^2) \partial^2 / \partial t^2 - \nabla^2)$ is itself a universal scalar, we may separate out the scalar and vector parts of the total equation to give *the wave equations*:

$$\square \phi = \rho / \varepsilon_0$$

and

$$\square \mathbf{A} = \mathbf{j} / \varepsilon_0$$

It is significant that $\phi$ and $\mathbf{A}$ are arbitrary to the point where they satisfy these equations (the condition of gauge invariance, as previously discussed under nonconservation). For convenience, we can arbitrarily restrict the values using a *gauge condition*. If we choose the so-called Lorentz gauge, in which

$$\frac{\partial \phi}{\partial t} + \nabla \cdot \mathbf{A} = 0,$$

and define new vectors $\mathbf{E}$ and $\mathbf{B}$, without reference to physical characteristics, such that

$$\mathbf{E} = -\nabla \phi + \frac{\partial \mathbf{A}}{\partial t}$$

and

$$\mathbf{B} = \nabla \times \mathbf{A},$$

we obtain the four Maxwell equations in their standard form, and identify $\mathbf{E}$ and $\mathbf{B}$ as the electrostatic and magnetic field vectors.

## 10 Conservation laws and fundamental symmetries

The outline derivations of classical mechanics and electromagnetic theory show that the group structure of space, time, mass and charge has the power to derive conventional results in a relatively simplified form. Numerous new mathematical results can be generated by even more direct uses of the symmetries. As previously noted, Noether's theorem requires the translation symmetry of time to be linked to the conservation of energy. Of course, since energy is related to mass by the equation $E = mc^2$, then the translation symmetry of time is also linked to the conservation of mass. To put it another way, the nonconservation of time is responsible for the conservation of mass. This is a result we could have derived from symmetry alone; and so, extending the analogy, we could link the conservation of the quantity of charge with the nonconservation, or translation symmetry of space; and since the latter is already linked with the conservation of linear momentum, we could propose a theorem in which the conservation of linear momentum was responsible for the conservation of the quantity of charge (of any type). By the same kind of reasoning, we could make the conservation of *type* of charge linked to the rotation symmetry of space, and so to the conservation of angular momentum, as in the following scheme:

| symmetry | conserved quantity | linked conservation |
| --- | --- | --- |
| space translation | linear momentum | value of charge |
| time translation | energy | value of mass |
| space rotation | angular momentum | type of charge |

In fact, we can immediately show these principles to be true in special cases. As Fritz London showed in 1927, the conservation of electric charge within a system is identical to invariance under transformations of the electrostatic potential by a constant representing changes of phase, and the phase changes are of the kind involved in the conservation of linear momentum. Since, in a conservative system, electrostatic potential varies only with the spatial coordinates, this is, in effect, a statement of the principle that the quantity of electric charge is conserved because the spatial coordinates are not, which is a special case of the first predicted relation.

In the second case, there is the relation between spin and statistics observed in fundamental particles. Fermions and bosons have different values of spin angular momentum; and they also differ in that fermions probably carry weak units of charge, where bosons have none. It thus appears to be the presence of a particular *type* of charge which determines the angular momentum state of the particle, so conservation of this type of charge is linked with the value of angular momentum. The validity of the full theorem, and its application to particle physics, can be derived using the quantum mechanical formalism derived in the next section.[11]

**11 The Dirac equation**

Yet another significant mathematical result follows from the basic representations of 4-vector space-time and quaternion mass-charge. A direct combination of these two constructs, putting the four parameters onto an equal overall footing in a single mathematical representation, produces a 32-part algebra which is identical in all respects to the 32-part algebra used in the Dirac equation for the electron, but much simpler in form and more powerful.[11-18] The components of this algebra can be described as:

| | |
|---|---|
| 2 complex numbers | (1,*i*) |
| 6 complex unit vectors | (1,*i*) × (**i**,**j**,**k**) |
| 6 complex unit quaternions | (1,*i*) × (***i***,***j***,***k***) |
| 18 complex vector quaternions | (1,*i*) × (**i**,**j**,**k**) × (***i***,***j***,***k***) |

Terms equivalent to the five gamma matrices (for example, ***ii***, ***ji***, ***ki***, ***j***, ***ik***) are easily derived.) A group version, with + and – units requires 64 terms (as does (S*, T*, M*, C*) × (S, T, M, C)).

Once the Dirac algebra has been established, it is a relatively easy process to show that the Dirac equation follows from quantization of a basic classical conservation equation, and the algebra, in this case, becomes simplified to a virtually pure quaternion algebra, as the vector element is removed via a scalar product. We begin with the Lorentz-invariant relationship between energy, mass and momentum:

$$E^2 - p^2 - m^2 = 0 \ .$$

Factorization of this expression requires the use of a complex and noncommutative algebra, viz. quaternions:

$$(kE + iip + ijm)(kE + iip + ijm) = 0 \ .$$

We can also incorporate the factor $e^{-i(Et - \mathbf{p} \cdot \mathbf{r})}$ without requiring new physical information:

$$(kE + iip + ijm)(kE + iip + ijm) \, e^{-i(Et - \mathbf{p} \cdot \mathbf{r})} = 0$$

In the classical equation, $E$ and $p$ are variables within the requirement that $E^2 - p^2$ is a constant for fixed $m^2$. However, quantization changes the status of these terms so that, for stationary quantum states, $E$ and $p$ become fixed, along with $m$. The variability now becomes confined to the space and time parameters incorporated into the exponential term, which can now be seen, physically, to represent the entire group of space and time translations and rotations.

A more general variability of space and time can be incorporated by replacing the factor (***k****E* + ***ii****p* + ***ij****m*) from the left with the differential operator

$$\left( i k \frac{\partial}{\partial t} + i \nabla + ijm \right)$$

acting on the 'wavefunction'

$$\psi = (kE + iip + ijm) \, e^{-i(Et - \mathbf{p} \cdot \mathbf{r})} \ ,$$

producing the expression

$$\left(i\mathbf{k}\frac{\partial}{\partial t} + \mathbf{i}\nabla + \mathbf{ij}m\right)\psi = 0,$$

which we recognise as the Dirac equation (in a form already second quantized because the quantization process has been applied to both the differential operator and the wavefunction).

It will be recognised that the need for quantization of $E$ and $p$ comes from the inverse relation between $m$ and $t$, and that the process has a profound effect on the classical energy-conservation equation. In quantizing via the Dirac equation, and, at the same time, imposing Lorentz-invariance, we effectively restructure the properties of the physical quantities involved, though our *physical* interpretation of the quantities remains unchanged.

In incorporating both the explicit quantization of *E-p-m* and the quaternion operators, the Dirac equation combines S* and T with an effective restructuring of M with the properties of C. Significantly, this has five components. The fact that only one direction of spin is well-defined is a consequence of using S* for S.

In classical relativistic theory, we emphasize the 4-vector nature of ($iE$, **p**) and describe $E^2 - p^2$ as an invariant, but, here we incorporate the invariance directly, and define a new term with five components, ($\mathbf{k}E + \mathbf{ii}p + \mathbf{ij}m$), with quantized rest mass. This 5-'dimensional' quantity combines the effects of 3-dimensional conserved and nonconserved parameters (the momentum term $p$ having 3 dimensions, although only one is normally defined). In effect, we structure mass (or energy-momentum-mass) as a 3-dimensional quantized and conserved parameter, like charge (with one of the 'dimensions' being itself dimensional). This is the result previously achieved by structuring charge-mass-space-time as a quaternion, with charge as the real or identity element.

## 12 Conclusion

The prediction of new mathematical theorems and the derivation of new algebraic concepts, as well as the procedures for obtaining standard theorems in classical mechanics, electromagnetic theory and quantum mechanics, show that the method of symmetry based on fundamental basic principles is not just a philosophical issue, but also a powerful method of generating new results, and of codifying existing ones. In fact, it is almost inevitable that new discoveries will follow after any successful exercise in getting down to the basics.

# References

1. G. J. Whitrow, *The Natural Philosophy of Time* (Nelson, London, 1961), pp. 135-57
2. P. Coveney and R. Highfield, *The Arrow of Time* (London, 1990), pp. 28, 143-4, 157.
3. A. Robinson, *Non-Standard Analysis* (Princeton University Press, 1996, original publication, 1966).
4. P. Rowlands, *Waves Versus Corpuscles: The Revolution That Never Was* (PD Publications, Liverpool, 1992)
5. P. Rowlands, Quantum indeterminacy, wave-particle duality and the physical interpretation of relativity theory from first principles. *Proceedings of Conference on Physical Interpretations of Relativity Theory III*, British Society for Philosophy of Science, London, September 1992, 296-310.
6. P. Rowlands, Quantum uncertainty, wave-particle duality and fundamental symmetries, in S. Jeffers, S. Roy, J-P. Vigier and G. Hunter (eds.), *The Present Status of the Quantum Theory of Light: A Symposium in Honour of Jean-Pierre Vigier* (*Fundamental Theories of Physics*, vol. 80, Kluwer Academic Publishers), 1997, 361-372.
7. P. Rowlands, The Fundamental Parameters of Physics. *Speculat. Sci. Tech.*, **6**, 69-80 (1983).
8. P. Rowlands, A new formal structure for deriving a physical interpretation of relativity. *Proceedings of Conference on Physical Interpretations of Relativity Theory II*, British Society for Philosophy of Science, London, September 1990, 264-8.
9. P. Rowlands, *The Fundamental Parameters of Physics: An Approach towards a Unified Theory* (PD Publications, Liverpool, 1991).
10. P. Rowlands and J. P. Cullerne, A Symmetry Principle for Deriving Particle Structures. *Proceedings of ANPA XX*, Cambridge, September 1998.
11. P. Rowlands and J. P. Cullerne, Applications of the nilpotent Dirac state vector (33 pp), xxx.lanl.gov/quant-ph/0103036.
12. P. Rowlands, An algebra combining vectors and quaternions: A comment on James D. Edmonds' paper. *Speculat. Sci. Tech.,* **17**, 279-282 (1994).
13. P. Rowlands, Some interpretations of the Dirac algebra. *Speculat. Sci. Tech.*, **19**, 243-51 (1996).
14. P. Rowlands, A New Algebra for Relativistic Quantum Mechanics. *Proceedings of Conference on Physical Interpretations of Relativity Theory V*, British Society for Philosophy of Science, London, September 1996, 381-7.
15. P. Rowlands, The physical consequences of a new version of the Dirac equation, in G. Hunter, S. Jeffers and J.-P. Vigier (eds.), *Causality and Locality in Modern Physics and Astronomy: Open Questions and Possible Solutions*, Kluwer, 1998, 397-402.
16. P. Rowlands, Further Considerations of the Dirac Algebra. *Proceedings of Conference on Physical Interpretations of Relativity Theory VI*, British Society for Philosophy of Science, London, September 1998.
17. P. Rowlands and J. P. Cullerne, The connection between the Han-Nambu quark theory, the Dirac equation and fundamental symmetries. *Nuclear Physics* **A 684**, 713-715 (2001).
18. P. Rowlands and J. P. Cullerne, The Dirac algebra and its physical interpretation (36 pp), xxx.lanl.gov/quant-ph/00010094.